\documentclass[aps,pra,showpacs,twocolumn,superscriptaddress]{revtex4}
\usepackage{amsmath,graphics}
\usepackage[next]{inputenc}
\usepackage[dvips]{epsfig}
\usepackage{hyperref}
\def\be{\begin{equation}}
\def\ee{\end{equation}}

\def\bi{\begin{itemize}}
\def\ei{\end{itemize}}
\def\bn{\begin{enumerate}}
\def\en{\end{enumerate}}
\def\bea{\begin{eqnarray}}
\def\eea{\end{eqnarray}}
\def\no{\nonumber}
\def\ba{\begin{array}}
\def\ea{\end{array}}
\def\bd{\begin{displaymath}}
\def\ed{\end{displaymath}}

\begin{document}
\title{Low energy states dynamic of entanglement for spin systems \footnote{This paper is dedicated to Prof. Y.
Sobouti the founder of Institute for Advanced Studies in Basic
Sciences}}

\author{R. Jafari}
\affiliation{Department of Physics, Institute for Advanced
Studies in Basic Sciences (IASBS), Zanjan 45137-66731, Iran}
\email[]{jafari@iasbs.ac.ir}

\begin{abstract}
We have composed the ideas of quantum renormalization group and
quantum information by exploring the low energy states dynamic of
entanglement resources of a system close to its quantum critical
point. We demonstrate the low energy states dynamical quantities
of the one dimensional magnetic systems could show the quantum
phase transition point and shows the scaling behavior in the
vicinity of the transition point. To present our idea, we study
the evolution of two spins entanglement in the one-dimensional
Ising model in the transverse field. The system is initialized as
the so-called thermal ground state of the pure Ising model. We
investigate evolvement of the generation of entanglement with
increasing the magnetic field. We have obtained that the
derivative of the time at which the entanglement reaches its
maximums with respect to the transverse field, diverges at the
critical point and its scaling behaviors versus the size of the
system are as same as the static ground state entanglement of the
the system.

\end{abstract}
\date{\today}


\maketitle
\section{Introduction \label{introduction}}
A fundamental difference between quantum and classical physics is
the possible existence of nonclassical correlations in quantum
systems. The physical property  responsible for this quantum
correlation is called Entanglement \cite{Bell}. Entanglement has
been recognized as an important resource for quantum information
and computation \cite{Nielsen}. However the role of entanglement
in quantum phase transition (QPT) \cite{Sachdev} is of
considerable interest. QPT as well as classical ones are
characterized by detecting nonanalytic behaviors in some physical
properties of the system. It is often accompanied by divergence in
some correlation functions, but quantum systems possess additional
correlations which do not exist in a classical counterpart, the
entanglement. Entanglement is a direct measure of quantum
correlations and shows nonanalytic behavior such as discontinuity
in the vicinity of the quantum phase transition point
\cite{Osterloh,Osborne}. An important motivation to study the
interconnection between condensed matter and quantum information
is to investigate whether it is possible to better characterize
condensed matter states by looking at their entanglement
properties. Recently, there has been extensive analysis of
entanglement in quantum spin models \cite{Amico1}. Various models
were considered for entanglement generation and their static
\cite{Osterloh} as well as dynamical properties \cite{Amico2}were
investigated. A thorough understanding of the dynamical evolution
of entanglement in the spin models has obviously implications for
the performance of quantum information processing, as well as for
understanding of fundamental quantum mechanics.

Our main purpose in this work is to compose the ideas of quantum
renormalization group \cite{Pfeuty1} and quantum information
theory to study the evolution of the dynamical properties of the
spin models in low energy states. To have a concrete discussion,
the one dimensional $S=\frac{1}{2}$ Ising model in transverse
field (ITF) has been considered by implementing the quantum
renormalization group (QRG) approach
\cite{miguel1,kargarian,Jafari1}.

\section{Quantum Renormalization Group \label{QRG}}

The main idea of the RG method is the mode elimination or thinning
of the degrees of freedom followed by an iteration which reduces
the number of variables step by step until reaching a fixed point.
In Kadanoff's approach, the lattice is divided into blocks. Each
block is treated independently to build the projection operator
onto the lower energy subspace. The projection of the inter-block
interaction is mapped to an effective Hamiltonian ($H^{eff}$)
which acts on the renormalized subspace \cite{miguel2,Langari,Jafari2}.


We have considered the ITF model
on a periodic chain of $N$ sites with Hamiltonian
\be
H=-J\sum_{i=1}^{N}(\sigma_{i}^{x}\sigma_{i+1}^{x}+g\sigma_{i}^{z}).
\label{eq1}
\ee

where $J>0$ is the exchange coupling and $g$ is the transverse
field. From the exact solution \cite{Pfeuty2} it is known that a
second order phase transition occurs for $g_{c}=1$ where
the behavior of the order parameter or magnetization is given
by $<\sigma^{x}>=(1-g)^{1/2}$ for $g<1$ and $<\sigma^{x}>=0$ for
$g>1$.

To implement QRG the Hamiltonian is divided to two-site blocks,
$H^{B}=\sum_{I=1}^{N/2}h_{I}^{B}$ with
$h_{I}^{B}=-J(\sigma_{1,I}^{x}\sigma_{2,I}^{x}+g\sigma_{1,I}^{z})$.
The remaining part of the Hamiltonian is included in the
inter-block part, $H^{BB}=-J\sum_{I=1}^{N/2}(\sigma_{2,I}^{x}\sigma_{1,I+1}^{x}+g\sigma_{2,I}^{z}).$
where $\sigma_{j,I}^{\alpha}$ refers to the $\alpha$-component of
the Pauli matrix at site $j$ of the block labeled by $I$. The
Hamiltonian of each block ($h_{I}^{B}$) is diagonalized exactly
and the projection operator ($P_{0}$) is
constructed from the two lowest eigenstates,
$P_{0}=|\psi_{0}\rangle\langle\psi_{0}|+|\psi_{1}\rangle\langle\psi_{1}|$,
where $|\psi_{0}\rangle$ is the ground state and
$|\psi_{1}\rangle$ is the first excited state.
In this respect the effective Hamiltonian
($H^{eff}=P_{0}[H^{B}+H^{BB}]P_{0}$) is
similar to the original one (Eq.(\ref{eq1})) replacing the couplings
with the following renormalized coupling constants.
\bea \label{eq2}
J'=J\frac{2q}{1+q^{2}},~q=g+\sqrt{g^{2}+1},
~g'=g^{2}.
\eea

\section{Reduced Density Matrix and Evolution of Concurrence\label{Concurrence}}

The entanglement is a local quantity which includes the global
properties of a system. Generally, the global properties of a
system enters the entanglement effectively by summing over the
whole degrees of freedom except the local one. In other words, a
system can be supposed of a single site and a heat bath (the rest
of system). It is supposed that the effect of a heat bath can be
replaced by an effective single site quantity, {\it the
entanglement}. The effective single site represents the long range
properties of the model and not the microscopic ones. Therefor we
can enter the global properties of the model to  entanglement (the
local quantity) using the renormalization group idea. In this
respect, we always think of a two site model which can be treated
exactly. However, the coupling constants of the two site model are
the effective ones which are given by the renormalization group
procedure. This can be used as an new method to calculate the low
energy states dynamic of entanglement in a large system.

The two site Hamiltonian of ITF model in the space spanned by
$\{|\uparrow\uparrow\rangle,|\uparrow\downarrow\rangle,|\downarrow\uparrow\rangle,
|\downarrow\downarrow\rangle\}$ ($|\uparrow\rangle$ and
$|\downarrow\rangle$ denote the eigenstates of $\sigma^{z}$), can
be expressed as

\bea
\label{eq4}
H=-J\left(
  \begin{array}{cccc}
    2g & 0 & 0 & 1 \\
  0 & 0 & 1 & 0 \\
  0 & 1 & 0 & 0 \\
  1 & 0 & 0 & -2g
  \end{array}
\right)
\eea

However the time evolution operator $U(t)=e^{-iHt}$ of the two site Hamiltonian (Eq.(\ref{eq4}))
has the following form

\bea
\label{eq5}
U(t)=J\left(
  \begin{array}{cccc}
    U_{11}(t) & 0 & 0 & U_{14}(t) \\
  0 & U_{22}(t) & U_{23}(t) & 0 \\
  0 & U_{32}(t) & U_{33}(t) & 0 \\
  U_{41}(t) & 0 & 0 & U_{44}(t)
  \end{array}
\right)
\eea

where

\bea
\no
U_{11}(t)&=&i\frac{2g}{\sqrt{1+4g^{2}}}\sin(Jt\sqrt{1+4g^{2}})+\cos(Jt\sqrt{1+4g^{2}}),\\
\no
U_{14}(t)&=&U_{41}(t)=\frac{i}{\sqrt{1+4g^{2}}}\sin(Jt\sqrt{1+4g^{2}}),\\
\no
U_{22}(t)&=&U_{33}(t)=\cos Jt,~U_{23}(t)=U_{32}(t)=i\sin Jt,\\
\no
U_{44}(t)&=&U^{\ast}_{11}.
\eea

The density matrix for the two sites system at time $t$ is
$\rho(t)=U(t)\rho(0)U^{\dag}$, where $\rho(0)$ is the the density
matrix of system at $t=0$ which is the thermal equilibrium of the
system. We choose  the thermal ground sate of the pure Ising model
(no transverse field applied)
$\rho(0)=\frac{1}{2}(|\sigma^{x},+\rangle\langle\sigma^{x},+|+|\sigma^{x},-\rangle\langle\sigma^{x},-|)$
as the initial state which the concurrence equals zero.

Here $|\sigma^{x},+\rangle$ and $|\sigma^{x},-\rangle$ are the two degenerate
ground state of Ising model with all spins pointing respectively to
the positive and negative $x$ direction.

\begin{figure}
\begin{center}
\includegraphics[width=9cm]{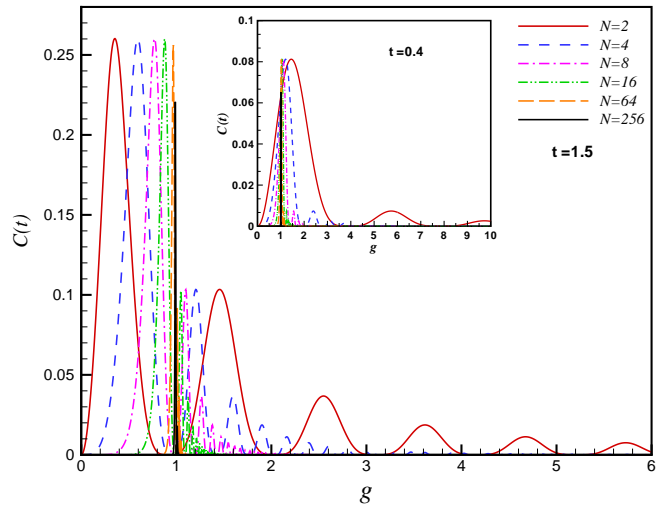}
\caption{(Color online) Evolution of the concurrence under RG
versus $g$ for $t=1.5$ and in the inset for $t=0.4$.} \label{fig1}
\end{center}
\end{figure}

For bipartite entanglement, a commonly used measure for arbitrary
state of two qubits is the so called concurrence \cite{Wootters}.
The concurrence is defined as

\bea
\label{eq6}
C(t)=max\{0,2\lambda_{max}(t)-tr\sqrt{\rho(t)\widetilde{\rho}(t)}\}
\eea

where $\widetilde{\rho}(t)=(\sigma^{y}\otimes\sigma^{y})\rho^{\ast}(t)(\sigma^{y}\otimes\sigma^{y})$, and
$\lambda_{max}$ is the largest eigenvalue of the matrix $\sqrt{\rho(t)\widetilde{\rho}(t)}$.

Therefor the analytic expression of the concurrence in terms of the parameters defined for the two site system is

\bea
\label{eq7}
C(t)=\frac{1}{2}\Big[1-\sqrt{1-(\frac{4g}{1+4g^{2}})^{2}\sin^{4}(\sqrt{1+4g^{2}}Jt)}\Big].
\eea

We have plotted the evolution of $C(t)$ under RG steps versus $g$
for $t=1.5$ and $t=0.4$ in Fig.\ref{fig1} which shows
that the concurrence changes from the equilibrium state and start
to oscillating when the external magnetic filed is turned on. As
$g$ increases, the height of each peak decreases gradually and
finally vanish as $g\rightarrow\infty$. Increasing the length of
chain enhances the oscillation of the concurrence versus the magnetic field.
However as the length of chain
increases the first peak of concurrence approaches the critical
point ($g_{c}=1$) and at the thermodynamic limit the system
becomes disentangled except at the critical point. Surviving of
the concurrence at the critical point is the results of the
correlation length divergence at $g_{c}=1$.

\begin{figure}
\begin{center}
\includegraphics[width=8.5cm]{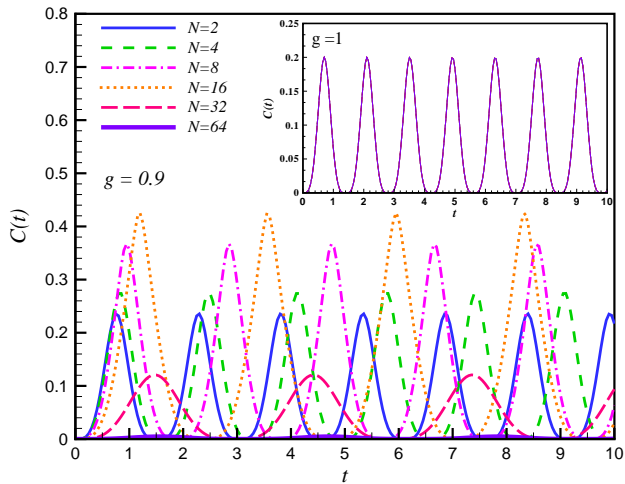}
\caption{(Color online) Concurrence of the ITF model as a function
of $t$ for different length chain for $g=0.9$. The inset shows the
concurrence of different lattice sizes collapse on a single curve
at the critical point ($g_{c}=1$).} \label{fig2}
\end{center}
\end{figure}

The evolution of concurrence under RG has been plotted in
Fig.\ref{fig2} versus $t$ for $g=0.9$ and $g_{c}=1$. From the
Eq.(\ref{eq7}), it is easy to see that the concurrence is
periodically fluctuating with the time $t$ with period of
$T=\frac{2\pi}{J\sqrt{1+4g^2}}$. Figure.\ref{fig2} shows that
the concurrence reduces under RG (increasing the size of system)
and disappears in the large system. But for $g_{c}=1$ there is no
concurrence reduction under RG and the concurrence of different
length chains coincide with each other (inset of Fig.
\ref{fig2}).

The non-analytic behaviour in some physical quantity is a feature
of second-order quantum phase transition. It is also accompanied
by a scaling behaviour since the correlation length diverges and
there is no characteristic length in the system at the critical
point. As we have stated in the RG approach for ITF model, a large
system, i.e. $N=2^{n+1}$, can be effectively describe by two sites
with the renormalized coupling of in the n-th RG step. Thus, the
concurrence between the two renormalized sites represents the
entanglement between two parts of the system each containing $N/2$
sites effectively. In this respect we can speak of {\it block
entanglement} -the entanglement between a block and the rest of
system- in a large system provided the size of the block and the
rest of system is equal.

For any $g$, there is a time $T_{max}^{k}(g)$ at which the $C(t)$
reaches its $k$th maximum (Fig.(\ref{fig2})) and is analyzed as a
function of coupling $g$ at different RG steps which manifest the
size of system. The first derivative of $T_{max}^{k}$ with respect
to the coupling constant ($\frac{dT_{max}^{k}}{dg}$) shows a
singular behavior at the critical point.

\begin{figure}
\begin{center}
\includegraphics[width=8.5cm]{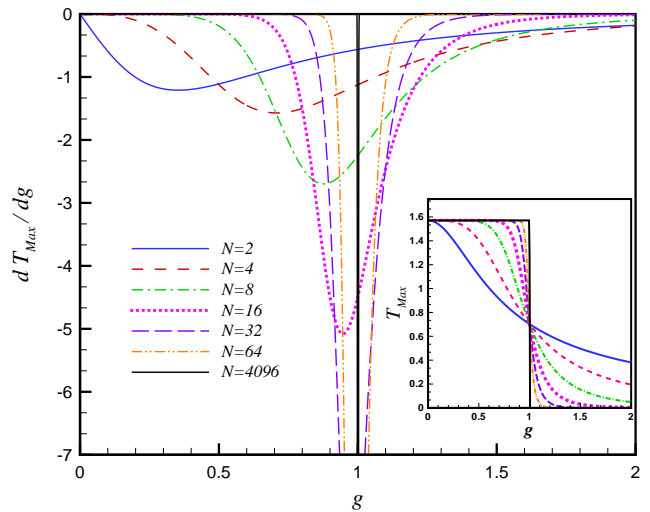}
\caption{(Color online) Evolution of the first derivative of
$T_{max}^{1}$ respect to magnetic field ($g$) under RG. The inset
shows the $T_{max}^{1}$ at different RG steps.} \label{fig3}
\end{center}
\end{figure}

\begin{figure}
\begin{center}
\includegraphics[width=9cm]{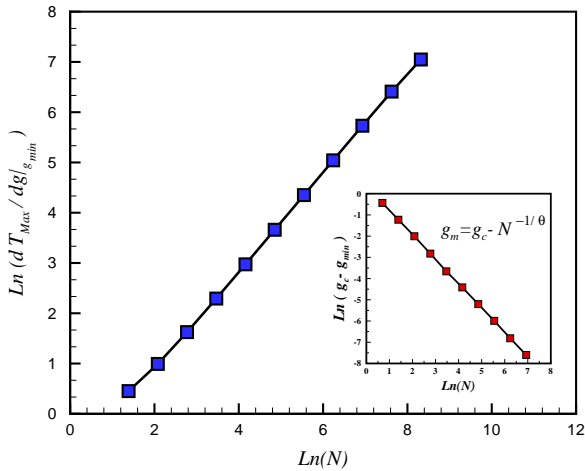}
\caption{(Color online) scaling the minimum of
$\frac{dT_{max}^{1}}{dg}$ for various size of system. The inset
shows the scaling of the position ($g_{m}$) of
$\frac{dT_{max}^{1}}{dg}$ for different length chains.}
\label{fig4}
\end{center}
\end{figure}

We have plotted $\frac{dT_{max}^{k}}{dg}$ for $k=1$ (first maximum
of the $C(t)$) versus $g$ in Fig.\ref{fig3} for different RG steps
which shows the singular behaviour as the size of system becomes
large. Surveying the detail shows that the position of the minimum
($g_m$) of  $\frac{dT_{max}^{k}}{dg}$ tends towards the critical
point like $g_{m}=g_{c}-N^{1/\theta}$ in which $\theta=1$ (inset
of figure \ref{fig4}). Moreover, we have derived the scaling
behavior of $\ln \frac{dT_{max}^{1}}{dg}|_{g_{m}}$ versus N. 
This has been plotted in Fig.(\ref{fig4}), which shows a linear
behavior of $\ln \frac{dT_{max}^{1}}{dg}|_{g_{m}}$ versus $ln(N)$.
The scaling behavior is $\ln \frac{dT_{max}^{1}}{dg}|_{g_{m}}\propto N^{\theta}$ with exponent
$\theta=1$.

It is easy to show that the exponent $\theta$ is directly related
to the correlation length exponent ($\nu$) close to the critical
point. The correlation length exponent, gives the behavior of
correlation length in the vicinity of $g_{c}$, i.e., $\xi\sim
(g-g_{c})^{-\nu}$. Under the RG transformation, Eq. (\ref{eq2}),
the correlation length scales in the $n$th RG step as
$\xi^{(n)}\sim (g_{n}-g_{c})^{-\nu}=\xi/n_{B}^{n}$ , which
immediately leads to an expression for $|\frac{dg_{n}}{dg}|_
{g_{c}}$ in terms of $\nu$ and $n_{B}$ (number of sites in each
block). Dividing the last equation to $\xi\sim (g-g_{c})^{-\nu}$
gives $|\frac{dg_{n}}{dg}|_{g_{c}}\sim N^{1/\nu}$ , which implies
$\theta=1/\nu$, since $\frac{dT_{max}^{K}}{dg}|_{g_{m}}\sim
|\frac{dg_{n}}{dg}|_{g_{c}}$ at the critical point. It should also
be noted that the scaling of the position of minimum, $g_{m}$
(inset of figure \ref{fig3}), also comes from the behavior of the
correlation length near the critical point. As the critical point
is approached and in the limit of large system size, the
correlation length almost covers the size of the system, i.e.,
$\xi\sim N$, and a simple comparison with $\xi\sim
(g-g_{c})^{-\nu}$ results in the following scaling form
$g_{m}=g_{c}-N^{1/\nu}$ .

\begin{figure}[t]
\begin{center}
\includegraphics[width=8.5cm]{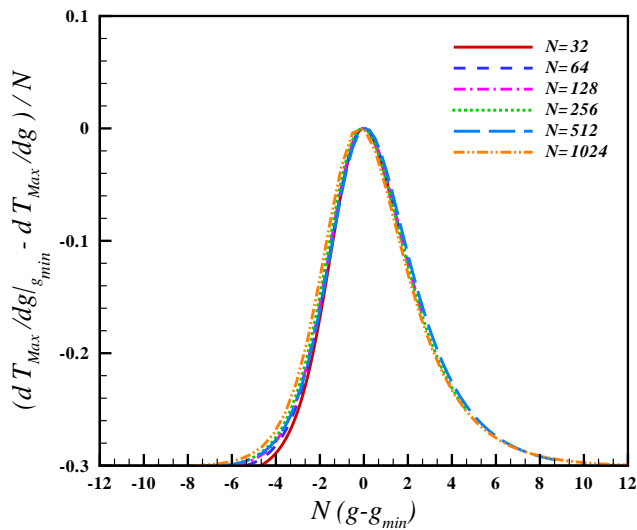}
\caption{(Color online) Finite-size scaling through the RG
treatment for different lattice sizes. The curves which correspond
to different system sizes clearly collapse on a single curve.}
\label{fig5}
\end{center}
\end{figure}

To obtain the finite-size scaling behavior of
$\frac{dT_{max}^{k}}{dg}|_{g_{m}}$, we look for a scaling function
in such away that all graphs tend to collapse on each other under
RG evolution which results in a large system. This is also a
manifestation of the existences of the finite size scaling for the
case of block entanglement. We have plotted
$\frac{dT_{max}^{k}}{dg}|_{g_{m}}-\frac{dT_{max}^{k}}{dg}$ versus
$N(g-g_m)$ for $k=1$ in Fig.\ref{fig5}. The upper curves which are for large
system sizes clearly show that all plots fall on each other.

The similar scaling behaviours and their relation to correlation
length exponent have been reported in our previous work
\cite{kargarian} in which we have studied the static properties of
the ground state entanglement of ITF model by RG method.

We would like to mention that in Ref.\cite{Chang} the authors
investigate the dynamics of concurrence of two nearest-neighbor
sites at ITF model using the exact solution. They shows that
$\frac{dT_{max}^{1}}{dg}$ dose not diverge at the critical point
but has a minimum at $g=1$. The divergence of
$\frac{dT_{max}^{1}}{dg}$ at the $g_{c}$ in our work and the
similarity of the dynamic scaling behaviors to the static scaling
behaviors originates from the low energy state properties.

\section{Summary and conclusions \label{conclusion}}

In this article, we have implement the idea of renormalization group
(RG) to study the low energy states dynamic of entanglement for
spin chains. In this respect we show that the RG procedure could
be implemented to obtain low energy states dynamic of systems in
terms of effective Hamiltonian which is described by renormalized
coupling constants. This manifest the fact that some dynamical
quantities of the system could show the fingerprint of quantum
phase transition for an infinite size system. These notions have
been observed and approved in our study of the ITF model.
Moreover, the RG approach shows that as the size of the system
becomes large, the derivative of the time at which the
entanglement reaches its maximums with respect to the transverse
field, diverges at the critical point and its scaling behaviors
versus the size of the system are as same as the static ground
state entanglement of the system.

\begin{acknowledgments}
The author would like to thank A. Langari, R. Fazio, A. G.
Moghaddam, Y. Sobouti, and V. Karimipour  for fruitful discussions
and comments.
\end{acknowledgments}

\end{document}